\documentclass[aps,prl,preprint,groupedaddress,showpacs]{revtex4-1}
\usepackage{graphicx}

\usepackage{color}
\usepackage{amsmath}
\definecolor{Blue}{rgb}{0.00, 0.00, 1.00}
\definecolor{Red}{rgb}{1.00, 0.00, 0.00}

\begin{document}

% \title{Record models for rainfall}

\title{An exactly solvable record model for rainfall}

\author{Satya N. Majumdar$^1$, Philipp von Bomhard$^2$ and Joachim Krug$^3$}

\affiliation{$^1$ Universit\'e Paris-Sud, CNRS, LPTMS, UMR 8626,
  91405 Orsay, France}
\affiliation{$^2$Deutsche R\"uckversicherung AG, Hansaallee 177, 40549
  D\"usseldorf, Germany}
\affiliation{$^3$Institute for Biological Physics, University of Cologne, 50937 K\"oln, Germany}

\date{\today}

\begin{abstract}
Daily precipitation time series are composed of null
entries corresponding to dry days and nonzero entries that describe
the rainfall amounts on wet days. Assuming that wet days
follow a Bernoulli process with success probability $p$, we show that the presence of dry
days induces negative correlations between record-breaking precipitation events. 
The resulting non-monotonic behavior of the Fano factor of the record
counting process is recovered in empirical data. We derive the full
probability distribution $P(R,n)$ of the number of records $R_n$ up to 
time $n$,
and show that for large $n$, its large deviation form coincides with
that of a Poisson distribution with parameter $\ln(p\,n)$.
% $P(R,n) \sim \exp\left[-\ln(p\,n)\, \Phi\left(\frac{R}{\ln (p\, 
% n)}\right)\right]$, 
% with an explicit rate function that 
% $\Phi(x)=1-x+x\ln (x)$.}}
We also study in detail the joint limit $p \to 0$, $n \to \infty$, 
which yields a random record model in continuous time $t = pn$. 
% Extending our analysis to alternating renewal models for dry and wet
% spells, we identify conditions on the tail exponent of the
% distribution of dry spells under which the number of records grows
% anomalously slowly.  

\end{abstract}


\pacs{}


%\keywords{}


\maketitle

An important and widely recognized consequence of global climate
change is an increase in the frequency of extreme weather conditions
such as heat waves, droughts and heavy precipitation
\cite{Easterling2000,Hansen2012,Coumou2012,Fischer2015,Cook2015}. 
The public perception of weather extremes is particularly sensitive to 
record-breaking events, which often receive extensive media
coverage. At the same time the analysis of records provides a useful
tool for the distribution-free inference of trends in time series,
because the temporal record statistics of sequences of independent 
random variables drawn from a continuous probability
distribution is manifestly universal
\cite{Foster1954,Glick1978,Arnold1998,Schmittmann1999,Benestad2004,Krug2007,Franke2012,Wergen2013,SM2014,Godreche2017}.
This observation has motivated a number of recent studies aimed at detecting and
quantifying the effects of a warming climate on the frequency of
temperature records \cite{Redner2006,Meehl2009,Wergen2010,Rahmstorf2011,Rowe2012,Coumou2013,Christiansen2013,Wergen2014}.   

In comparison, the effects of climatic trends on precipitation records
are more complex and have generally 
received less attention \cite{Benestad2003,Benestad2006,Lehmann2015}. 
In order to detect such trends, the null model describing a stationary
climate has to account for the specific structure of precipitation
time series. 
In contrast to temperature, which is well described as a continuous
random variable with a Gaussian distribution \cite{Wergen2010,Wergen2014}, the
amount of daily rainfall at a specific location has a positive
probability of being exactly zero. Stochastic precipitation models 
incorporate this basic feature by combining an \textit{occurrence process}
that determines whether a given day is dry (zero precipitation) or wet
(nonzero precipitation) with an \textit{amount process} that
specifies the amount of rainfall on a wet day \cite{Wilks1999}. 

In this Letter we show that the presence of dry days has a profound effect on the
occurrence statistics of precipitation records in a stationary climate. Assuming that
the wet days follow a Bernoulli process with success probability $p$,
we find that record events become negatively correlated when $p < 1$. 
This is in marked contrast to the well-known property of record events
from sequences of independent, identically and continuously
distributed (i.i.c.d.) random variables to be stochastically
independent \cite{Arnold1998,Wergen2013,SM2014,Godreche2017}.
As a consequence, the ratio of the variance and the mean of the
record counting process, known as the \textit{Fano factor}, displays a
minimum at intermediate times when $q=1-p$ is sufficiently large. This minimum is an unequivocal signature of correlations
between record events, and we demonstrate that it can be clearly
identified in empirical data. For this comparison we use time series 
comprising rainfall amounts on a given calendar day over several
decades, which justifies the assumption of uncorrelated occurrence and
amount processes. We expect that the mechanism giving rise to correlations in
the Bernoulli model is of broader relevance also beyond the specific
context of precipitation records, and provide a detailed
analysis of the model including the full distribution of the number of
records. 

\noindent
\textit{Bernoulli model.} Within the Bernoulli model a dry day with zero precipitation occurs with probability
$q$, and a wet day with probability $p=1-q$. For a wet day, the amount
of precipitation $x$ is a random variable
drawn from a continuous probability density $p_W(x)$ with support on
the positive real axis. The full probability density of precipitation
$x_n$ on day $n$ thus reads
\begin{equation}
p(x)= q\, \delta(x) +(1-q)\, p_W(x)\, .
\label{effectivepdf.1}
\end{equation}
The $\delta$-function at $x=0$ implies that the corresponding cumulative distribution function 
\begin{equation}
P(x)= \int_0^x dx' \, p(x')  = q\, \theta(x) +(1-q)\, P_W(x)\, 
\label{effectivepdf.2}
\end{equation}
is discontinuous at the origin, as indicated by the Heaviside theta
function. 
We are interested in the statistics of the number of
record events $R_n$ that have occurred up to time $n$. It is convenient to introduce a
binary indicator variable $\sigma_m$ for the $m$-th day such that $\sigma_m=1$ if a record occurs on the
$m$-th day, and $\sigma_m=0$ otherwise. Clearly
\begin{equation}
R_n= \sum_{m=1}^n \sigma_m\, .
\label{record_number.1}
\end{equation}
We note one important point: If a record occurs on the $m$-th day, then the $m$-th day is necessarily wet.

The mean number of records is given by
\begin{equation}
\langle R_n \rangle= \sum_{m=1}^n \langle \sigma_m \rangle = \sum_{m=1}^n r_m
\label{mean.11}
\end{equation}
where the record rate $r_m$ denotes the 
probability that a record occurs on the $m$-th day. The latter is given by
\begin{equation}
r_m= (1-q) \int_0^{\infty} dx\, p_W(x)\, P(x)^{m-1}\, ,
\label{mean.12}
\end{equation}
with the following intepretation: The probability that the $m$-th day is a wet day 
with precipitation $x>0$ is $(1-q)\,p_W(x)$, and in order for this to be
a record all the previous $(m-1)$ days must have precipation less than $x$. 
To perform the integral we make the substitution $x \to u = P(x)$,
noting that $u \in [q,1]$ and $du= (1-q)\, p_W(x) dx$ for $x > 0$. The
resulting expression
\begin{equation}
r_m= \int_q^1 du\, u^{m-1} = \frac{1-q^m}{m} \, 
\label{mean.15}
\end{equation}
is independent of the distribution $p_W(x)$ and reduces to the
classic result $r_m = 1/m$ for i.i.c.d. random variables when $q \to
0$. Correspondingly, the mean number of records up to day $n$ is given by 
\begin{equation}
\langle R_n \rangle = \sum_{m=1}^n \frac{1-q^m}{m} \, .
\label{mean.16}
\end{equation}
For large $n$ and fixed $q=1-p$, it is easy to show that $\langle 
R_n\rangle 
\approx \ln(p\,n)+ \gamma_E$, where $\gamma_E= 0.57721...$ is the Euler 
constant (see \cite{SM} for details). Thus at late times the record sequence 
looks like
a `diluted' i.i.c.d. record process where the effective number of random
variables that have been presented up to time $n$ is reduced by a
factor $p$. We will see below that this observation applies also to
the variance as well as to the full distribution of $R_n$. 

To compute the second moment of $R_n$, we square and average
Eq.~(\ref{record_number.1}), using that $\sigma_m^2=\sigma_m$. This gives
\begin{equation}
\langle R_n^2 \rangle = \langle R_n \rangle + 2 \sum_{l_1=1}^{n-1} \sum_{l_2=1}^{n-l_1} \langle \sigma_{l_1}\, \sigma_{l_1+l_2}\rangle,
\label{var.12}
\end{equation}
where $\langle \sigma_{l_1}\, \sigma_{l_1+l_2}\rangle$ is the joint
probability of two records occurring on day $l_1$ and $l_1+l_2$. To compute this, let the record at day $l_1$ have value $x_1$ and the one at $l_1+l_2$ have value $x_2$ with
$x_2>x_1$. Evidently, both days have to be necessarily wet. All the days before $l_1$ must have precipitation 
values less than $x_1$, and all the days between $l_1$ and $l_1+l_2$
must have precipitation values less than $x_2$. Writing down the
corresponding probability in analogy to Eq.~(\ref{mean.12}) and performing
the substitution $x \to P(x)$ (see \cite{SM}) leads 
to the simple form
\begin{equation}
\langle \sigma_{l_1}\, \sigma_{l_1+l_2}\rangle = \int_q^{1} du_2
\int_{q}^{u_2}\, du_1 \, u_1^{l_1-1}\, u_2^{l_2-1}\, 
\label{var.16}
\end{equation}
with $l_2 \geq 1$. For $l_2=0$, $\langle \sigma^2_{l_1}\rangle =r_{l_1}$.
Combining Eq. (\ref{var.16}) with the result (\ref{mean.15}) for the 
record rate yields the
connected correlation function of record events,
\begin{widetext}
\begin{equation}
g_{l_1,l_1+l_2} \equiv \langle \sigma_{l_1}\, \sigma_{l_1+l_2}\rangle - r_{l_1} r_{l_1+l_2} =
- \frac{q^{l_1}}{l_1} \int_q^1 du \, u^{l_2-1} (1 - u^{l_1}) =
- \frac{q^{l_1}}{l_1} \left( \frac{1-q^{l_2}}{l_2} - \frac{1-q^{l_1
  + l_2}}{l_1 + l_2} \right) 
\label{corr}
\end{equation}
\end{widetext}
which is universal (independent of $p_W(x)$) for all $l_1\ge 1$ 
and $l_2\ge 1$.
The second equality in Eq.~(\ref{corr}) manifestly shows that the correlation is
negative for all $l_1, l_2$ and $0 < q < 1$. Thus the record
events become anticorrelated when $q > 0$. The origin of these correlations ultimately lies in the discontinuity of
the distribution function (\ref{effectivepdf.2}), which reduces the domain of
integration in Eqs.~(\ref{mean.15}) and (\ref{var.16}) compared to the
i.i.c.d. case. We are however not aware of any intuitive explanation
for why the correlations are negative. Moreover, for fixed $l_1$ and
large $l_2$ , the connected
correlation function decays as a power law, $g_{l_1,l_1+l_2} \sim - q^{l_1}/[l_2(l_1+l_2)]\sim l_2^{-2}$.
This indicates that the record breaking events are rather strongly correlated.

Inserting Eq. (\ref{var.16})
into Eq. (\ref{var.12}) and performing the double sum yields, after a
substantial amount of algebra (see \cite{SM} for details), the expression
\begin{eqnarray}
V_n(q) =  \langle R_n^2\rangle- \langle R_n\rangle^2 =  \langle
                                                          R_n\rangle +
2\int_q^1
\frac{du \, u^n}{1-u} \left[ \int_q^{u} dv \frac{1-v^n}{1-v}-
\int_{q/u}^1 dv \frac{1-v^n}{1-v}\right] 
\label{var.25}
\end{eqnarray} 
for the variance of the number of records up to day $n$.
Asymptotically for large $n$ with fixed $q=1-p$, it can be shown~\cite{SM} that
$V_n(q)\to \langle R_n\rangle - \pi^2/6 \approx \ln(p\, n)+\gamma_E -
\pi^2/6$. 

\noindent
\textit{Random record model.} In order to arrive at a more tractable expression
for $V_n$, we now analyze the problem in the scaling limit $p \to 0$,
$n \to \infty$ at fixed $t = pn$. In this limit the Bernoulli sequence of
wet days becomes a Poisson process of unit intensity in continuous
time $t$. In the mathematical literature this setting is known as the
random record model \cite{Arnold1998,Westcott1977,Bunge1992}, see also
\cite{Davidsen2008,Park2008}. For the mean number of records (\ref{mean.16}) the limit $q \to 1$, $n \to
\infty$ yields $\langle R_n \rangle \to \mu(pn)$ with 
\begin{equation}
\mu(t)=\int_0^t dy\, \frac{1-e^{-y}}{y}= \ln t + \gamma_E +\int_t^{\infty} 
\frac{e^{-z}}{z}\, dz\, . 
\label{mean.17}
\end{equation}
The asymptotic behaviors of $\mu(t)$ are
$\mu(t)  \to   t- t^2/4$ as  $t\to 0$ and
$\mu(t) \to \ln t + \gamma_E$ as $t\to \infty$.
Thus, the scaling function describes a crossover in the mean number of 
records from an early time 
linear growth $\langle R_n \rangle \approx p\, n$ where the number of
records is limited by the number of events, 
to a late time logarithmic growth $\langle R_n \rangle \approx \ln(p\, n) +
\gamma_E$. 
Taking the scaling limit of the expression (\ref{var.25}) is not
straightforward, but eventually leads to the relatively simple form (see \cite{SM}) 
% \begin{widetext}
\begin{equation}
V_n(q) \to  \mu(t) + 2 \int_0^t \frac{dz}{z} e^{-z}\left[ \mu(t)- 
\mu(z)- \mu(t-z)\right]
\label{var.28}
\end{equation}
% \end{widetext}
where $\mu(t)$ is given in Eq. (\ref{mean.17}). 
 
\noindent
\textit{Fano factor.} To quantify the correlations between record events, it is useful
to introduce the Fano factor \cite{Fano1947} defined as the ratio of the variance to
the mean of the record counting process, 
$
F_n= \frac{V_n}{\langle R_n\rangle}
$. We first prove that $F_n$, for an arbitrary time-series, must 
be an 
increasing function of $n$ if
record events are uncorrelated. Let
$\langle \sigma_m\rangle=r_m$ denote the record rate 
at step $m$ of the time-series. In the 
absence of correlations between record events,
$\langle \sigma_l \sigma_m \rangle = r_m \delta_{l,m} + r_l r_m (1 -
\delta_{l,m})$, which implies using (\ref{var.12}) that
\begin{equation}
\label{Vuncorr}
V_n = \sum_{m=1}^n r_m (1-r_m).
\end{equation}
As a consequence
\begin{equation}
F_{n+1}-F_n = \frac{S_n}{\langle R_n \rangle} - \frac{S_{n+1}}{\langle
  R_{n+1} \rangle},
\end{equation}
where $S_n = \sum_{m=1}^n r_m^2$. Based on this relation it is easy to
show that $F_{n+1} - F_n > 0$ provided $r_{n+1} < r_m$ for all $m \leq n$,
which only requires the record rate to be monotonically
decreasing. Thus a non-monotonic behavior of $F_n$ is an unambiguous
signature of correlations. 

\begin{figure}[ht]\includegraphics[width=0.7\textwidth]{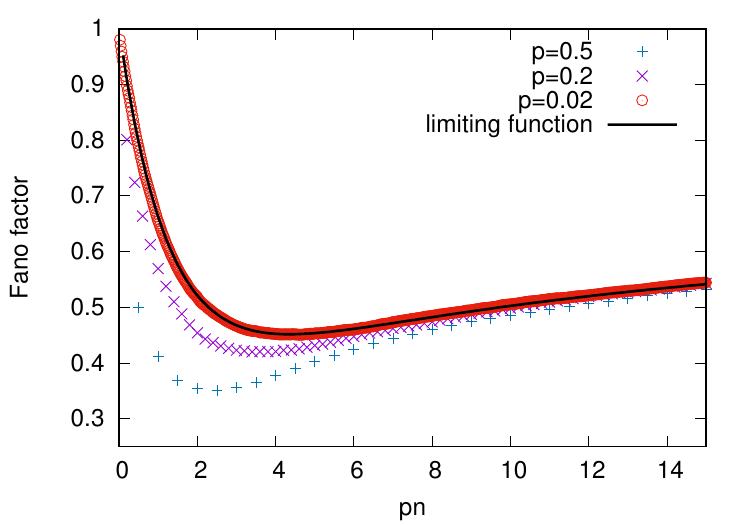}
%\begin{center}\includegraphics[width=0.241\textwidth]{rec_number43.png}\includegraphics[width=0.24\textwidth]{bos43.png}\end{center}
\caption{(Color online) The Fano factor of the record process obtained
  from simulations (symbols) is compared to the analytic limit
  function $F(t)$ in Eq.~(\ref{ff.2}) (full line). Note that the
  numerical estimates start at $F_1 = 1-p$. \label{fig.fano}} \end{figure}

Using the results from Eqs.~(\ref{mean.17}) and (\ref{var.28}), we find that
in the scaling limit the Fano factor converges to the scaling form, $F_n(q)\to F(t=p\, n)$ with
\begin{equation}
F(t) = \nonumber \\ 
1 +  \frac{2}{\mu(t)} \int_0^t \frac{dz}{z}\, e^{-z}\left[ \mu(t)-
\mu(z)- \mu(t-z)\right]\,.
\label{ff.2}
\end{equation}
The scaling function $F(t)$ is clearly
non-monotonic, showing that the strong correlations between record
events persist in the scaling limit (Fig.~\ref{fig.fano}).
It starts at $F(0)=1$, decreases with increasing $t$, reaches
a minimum around $t^* \approx 4.4$, and  
converges slowly back to $F=1$ as $t\to \infty$. Its asymptotic behaviors
can be easily computed from the exact expresssion in Eq.~(\ref{ff.2}),
and we obtain
$F(t) \to 1- t/2 + O(t^2)$ as $t \to 0$ and 
$F(t) \to 1- \pi^2/(6\, \ln t)$ as $t \to \infty$.
The figure also shows estimates for $F_n$ at finite $p > 0$ obtained
from simulations. It can be seen that the minimum is even more pronounced
at positive $p$, and the simulation results are
indistinguishable from the asymptotic prediction (\ref{ff.2}) for $p =
0.02$.

\noindent
\textit{Comparison to precipitation data.} In order to test the
predictions of the Bernoulli model we analyzed a large set of
daily precipitation data compiled by the German
weather service (DWD). The full data set comprises rainfall amounts
from 5400 weather stations positioned throughout Germany. Out of these, 417
stations were selected which provided complete daily precipitation
time series for the period 1974-2013 \cite{Bomhard2014}. The average
rainfall probability for this data set is close to $p=0.5$ with some
variability between stations. In order to minimize the effects of the
variability in $p$, we further restricted the analysis to those 
stations where the time-averaged precipitation probability lies in the
interval $p \in [0.48,0.52]$. This leaves 144 stations covering the 40
year period.
%from 1974 to 2013. 
For each station we extracted 365 time series corresponding to 
precipitation amounts on a given calendar day.

\begin{figure}[t]\includegraphics[width=0.7\textwidth]{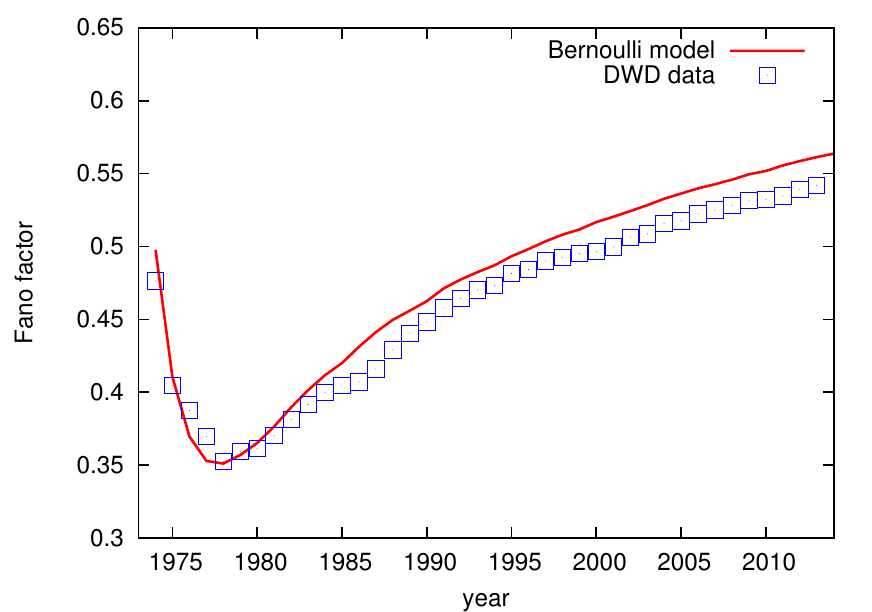}
%\begin{center}\includegraphics[width=0.241\textwidth]{rec_number43.png}\includegraphics[width=0.24\textwidth]{bos43.png}\end{center}
\caption{(Color online) Blue squares show the Fano factor of precipitation records estimated
from daily rainfall amounts at 144 German weather stations. For
comparison, the full line shows simulation results obtained from the Bernoulli model with the
average rainfall probability $p=0.5$. \label{fig.data}} \end{figure}

Figure \ref{fig.data} shows the Fano factor of the number of
precipitation records obtained from the empirical data, compared to
simulations of the Bernoulli model with $p=0.5$. The simulation data
were averaged over $5 \times 10^4$ runs, which is close to the total
number of empirical time series ($144 \times 365 = 52560$). We have
checked that allowing $p$ to vary over the interval $[0.48,0.52]$ in
the simulations does not significantly affect the results. The
empirically determined Fano factor displays a pronounced minimum and
the overall shape is in good agreement with the model. The remaining
discrepancy at longer times is probably not of a statistical nature
and could be related to features that are ignored in the model, such as spatial 
correlations between weather stations or trends in the model
parameters. 

\noindent
\textit{Distribution of the number of records.} Having derived the 
mean and the variance of the record number $R_n$, one may
naturally investigate its full distribution 
$P(R,n)={\rm Prob.}[R_n=R]$. Exploiting the renewal structure of the 
record process in the Bernoulli model, we were able to derive
a compact exact expression for the double generating function (see \cite{SM})
\begin{equation}
\sum_{n=0}^{\infty} \sum_{R=0}^{\infty} P(R,n) \lambda^R\, z^n = 
\frac{{(1-qz)}^{\lambda-1}}{(1-z)^\lambda}\,.
\label{dist.2}
\end{equation}
For $q=0$, the right hand side reduces to $(1-z)^{-\lambda}$, 
a known result for the i.i.c.d 
case~\cite{Renyi,Goldie,Flajolet-Sedgewick,GL2008,MMS2009}. From
Eq.~(\ref{dist.2}), one can in principle compute all the moments. 
Moreover, by analysing Eq. (\ref{dist.2}) for large $n$ 
(with fixed $p=1-q$ and $R\ge 1$), we can
show (see~\cite{SM}) that $P(R,n)$ converges
to the Poisson distribution
\begin{equation}
P(R,n) \approx \frac{1}{pn}\, \frac{\left(\ln (pn)\right)^{R-1}}{(R-1)!}\, .
\label{Poisson.1}
\end{equation}
We conclude that the record occurrence events
become a Poisson process in `time' $\ln(pn)$ for large $n$, as was   
observed previously for the i.i.c.d. case $q=0$ \cite{Sibani1998,GL2008}.

Interestingly, in the limit 
$R\to \infty$, $n\to 
\infty$, 
but keeping 
the ratio
$x= R/\ln (pn)$ fixed, the Poisson distribution in Eq. (\ref{Poisson.1}) admits a large 
deviation form
\begin{equation}
P(R,n) \sim e^{- \ln(p\, n)\Phi\left(\frac{R}{\ln (p\, n)}\right)}
\label{ldf.1}
\end{equation}
with an explicit rate function 
\begin{equation}
\Phi(x)= 1-x + x\, \ln x \, ; \quad x\ge 0 \, .
\label{ldf.2}
\end{equation}
Let us remark that the large deviation form in Eq. (\ref{ldf.1}) may look
a bit unfamiliar. Typically in statistical physics problems one finds a 
large deviation principle of the form
$\sim \exp[- L\, \Phi\left(\frac{R}{L}\right)]$, where $L$ represents the
`size' of the system. In the present problem, the effective size $L$ is not
$n$, but rather the average number of records $\langle R_n\rangle \sim
\ln (p\, n)$. Similar `anomalous' large deviation forms
appeared before in the context of the distribution of the number 
of zero crossings of smooth Gaussian fields in a certain time 
interval (or equivalently in the distribution of the
number of real roots of a class of random polynomials of degree 
$n$)~\cite{SM2007,SM2008,PS2018}, and more 
recently in the distribution of entanglement in random quantum spin 
chains~\cite{DMH2017}.

The rate function (\ref{ldf.2}) is independent of $q$. 
Typical fluctuations of $R_n$ are described by the quadratic
approximation of the rate function $\Phi(x)$ 
around its unique minimum at $x^*=1$.
Substituting this quadratic form in Eq. (\ref{ldf.1}), we find that
the typical fluctuations are described by a Gaussian 
with mean and variance $\ln (p\, n)$.
Thus despite the power law correlations between the indicator variables
$\sigma_m$,  
their sum $R_n= \sum_{m=1}^n \sigma_m$
satisfies a central limit theorem. 

\noindent
\textit{Conclusions.} Motivated by the statistics of rainfall,
we have investigated a simple extension of the classic
i.i.c.d. record problem where the non-negative random variables forming the time
series take on the value zero with a positive probability $q > 0$. Our
key finding is that this induces long-ranged correlations between
record events, which lead to a pronounced minimum in the Fano factor
of the record counting process. The emergence of correlations between
record events has been observed previously, e.g., for records drawn from distributions that
broaden \cite{Krug2007} or shift \cite{Franke2012,Wergen2011} in time,
or as a consequence of rounding effects \cite{Wergen2012}. Taken
together, these results highlight the fact that the stochastic
independence between record events in the standard i.i.c.d. setting is
a highly non-generic and fragile feature.

The comparison with the empirical data in Fig.~\ref{fig.data} shows that the
Bernoulli model qualifies as a null model for 
precipitation time series comprising daily rainfall amounts on a given
calendar day over a sequence of years. However, the model clearly fails to
describe time series of rainfall amounts on consecutive days, which
are characterized by strongly correlated spells of dry and wet days.   
This kind of data can be modeled by an alternating renewal process,
where dry and wet spell lengths are drawn independently from two
different probability distributions \cite{Wilks1999}. The record occurrence
statistics is then again universal with respect to the amount
distribution $p_W(x)$ but depends explicitly on the spell length
distributions. Detailed results for this model will be
reported elsewhere, focusing in particular on the
consequences of heavy-tailed distributions of dry spells \cite{MvBK}. 

\begin{acknowledgments}

We acknowledge the kind hospitality of MPI-KS Dresden, where this work
was initiated during the workshop \textit{Climate
  Fluctuations and Nonequilibrium Statistical Mechanics}.

\end{acknowledgments}

\newpage

\appendix

\setcounter{equation}{0}
\setcounter{figure}{0}
\renewcommand{\theequation}{S\arabic{equation}}
\renewcommand{\thefigure}{S\arabic{figure}}

\large{Supplemental material}

\normalsize

\section{Asymptotic behavior of the average number of records}

The average number of records up to time $n$ in the Bernoulli model is given in Eq. (7)
of the main text that reads
\begin{equation}
\langle R_n\rangle = \sum_{m=1}^n \frac{1-q^m}{m}\, .
\label{S1.avg1}
\end{equation}
To find the leading asymptotic behavior for large $n$ and any $0\le q\le 1$, we first 
note that as $n\to \infty$
\begin{equation}
\sum_{m=1}^n \frac{1}{m}= \ln (n) + \gamma_E + O\left(\frac{1}{n}\right)
\label{S1.avg2}
\end{equation}
where $\gamma_E=0.57721\ldots$ is the Euler constant. Furthermore, for $0\le q\le 1$ and 
$n\to \infty$ we have
\begin{equation}
\sum_{m=1}^n \frac{q^m}{m}= \sum_{m=1}^{\infty} \frac{q^m}{m}- \sum_{m=n+1}^{\infty} 
\frac{q^m}{m}= -\ln(1-q) + O\left(q^{n+1}\right)\, .
\label{S1.avg3}
\end{equation}
Subtracting Eq. (\ref{S1.avg3}) from (\ref{S1.avg2}), one gets using $p=1-q$, the 
following asymptotic behavior as $n\to \infty$
\begin{equation}
\langle R_n\rangle = \ln(p\, n) + \gamma_E + O\left(\frac{1}{n}\right)
\label{S1.avg4}
\end{equation}
as announced after Eq. (7) of the main text.

\section{Derivation of the Variance of $R_n$}

In this section, we provide a derivation of the main results for the variance in Eqs. (11) and (13) of the main text.
On the way, we also give a derivation of Eq. (9) of the main text.
We start from Eq. (8) of the main text that reads
\begin{eqnarray}
\langle R_n^2 \rangle &= & \langle R_n \rangle + 
2 \sum_{m_1<m_2} \langle \sigma_{m_1}\, \sigma_{m_2}\rangle \nonumber \\
& =& \langle R_n \rangle + 
2 \sum_{l_1=1}^{n-1} \sum_{l_2=1}^{n-l_1} \langle \sigma_{l_1}\, \sigma_{l_1+l_2}\rangle
\label{S1.1}
\end{eqnarray}
where $\langle R_n\rangle$ is given by Eq. (\ref{S1.avg1}).
To compute the correlation function $\langle \sigma_{l_1}\, \sigma_{l_1+l_2}\rangle$, we note that it is simply
\begin{equation}
\langle \sigma_{l_1}\, \sigma_{l_1+l_2}\rangle = {\rm Prob.}\left[
  {\rm {a\,\, record\,\, happens\,\, at\,\, day}} \,\, l_1 \,\,
{\rm {and\,\,
a\,\, record\,\, happens\,\, at\,\, day}}\,\, l_1+l_2 \right] \, .
\label{S1.2}
\end{equation}
To compute this joint probability, let the record at day $l_1$ have value $x_1$ and the one at $l_1+l_2$ value $x_2$ with
$x_2>x_1$. Evidently, both days have to be necessarily wet. All the days before $l_1$ must have precipitation
values less than $x_1$.
In addition, all the days between $l_1$ and $l_1+l_2$ must have precipitation values less than $x_2$ (this is needed
if $x_2$ is a record). Hence, using the independence of days and
knowing that the $l_1$-th day and the $l_1+l_2$-th days
are necessarily wet, we get
\begin{equation}
\langle \sigma_{l_1}\, \sigma_{l_1+l_2}\rangle =  \int_\delta^\infty dx_2 \int_{\delta}^{x_2} dx_1 \left[(1-q)\, p_W(x_2)\right]\,
\left[\int_0^{x_2}
p(x')\, dx'\right]^{l_2-1} \, \left[(1-q)\,p_W(x_1)\right]\, \left[\int_0^{x_1} p(x')\, dx'\right]^{l_1-1}\, ,
\label{S1.3}
\end{equation}
where $p(x)=q \delta(x)+ (1-q) p_W(x)$ is given in Eq. (1) of the main text. 
Note that we have introduced, for convenience, a lower cut-off $\delta$
in the integrals over $x_1$ and $x_2$. This is to indicate that a record
occurs only on a wet day where the precipitation is strictly positive,
i.e., the distibution $p_W(x)$ has support only over $x\in 
[\delta,\infty]$ with $\delta\to 0^+$. Thus we will keep this cut-off 
$\delta$ in the $x$-integrals and eventually take the limit $\delta\to 
0^+$.

To proceed further, we
make the change of variable
\begin{equation}
u= \int_0^x p(x') dx' = q\, \theta(x) + (1-q) \int_\delta^x p_W(x')\, dx' \, .
\label{S1.4}
\end{equation}
Consequently, the complicated integral in Eq. (\ref{S1.3}),
upon taking $\delta\to 0^+$ limit, simplifies 
nicely to yield
\begin{equation}
\langle \sigma_{l_1}\, \sigma_{l_1+l_2}\rangle = 
\int_q^{1} du_2 \int_{q}^{u_2}\, du_1 u_1^{l_1-1}\, u_2^{l_2-1}
\label{S1.5}
\end{equation}
valid for all $l_1\ge 1$ and $l_2\ge 1$.
This then provides a derivation of Eq. (9) in the main text.
As in the case of the mean $r_m= \langle \sigma_m\rangle$ in Eq. (6) of the main text, this two point correlation
function is also universal, i.e., independent of $p_W(x)$.

Plugging Eq. (\ref{S1.5}) into Eq. (\ref{S1.1}) gives
\begin{equation}
\langle R_n^2 \rangle =  \langle R_n \rangle + 2  \sum_{l_1=1}^{n-1} \sum_{l_2=1}^{n-l_1} \int_q^{1} du_2\, u_2^{l_2-1}
\int_q^{u_2} du_1\, u_1^{l_1-1} \, .
\label{S1.6}
\end{equation}
We first perform the sum over $l_2$ which is a simple geometric series and obtain
\begin{equation}
\langle R_n^2 \rangle =  \langle R_n \rangle + 2  \sum_{l_1=1}^{n-1}\int_q^1 du_2 \,
\left[\frac{1-u_2^{n-l_1}}{1-u_2}\right]\, \int_q^{u_2} du_1\, u_1^{l_1-1}\, .
\label{S1.7}
\end{equation}
Next we note that the sum over $l_1$ from $1$ to $n-1$ can be extended up to $l_1=n$, since the $l_1=n$ term
is identically $0$. This step turns out to be rather convenient. Hence
\begin{equation}
\langle R_n^2 \rangle =  \langle R_n \rangle + 2  \sum_{l_1=1}^{n}\int_q^1 du_2 \,
\left[\frac{1-u_2^{n-l_1}}{1-u_2}\right]\, \int_q^{u_2} du_1\, u_1^{l_1-1}\, .
\label{S1.8}
\end{equation}
Finally, performing the geometric sum over $l_1$ gives
\begin{eqnarray}
\langle R_n^2 \rangle &= & \langle R_n \rangle + 2
\int_q^1 \frac{du_2}{1-u_2} \int_q^{u_2} du_1
\left[\frac{1-u_1^n}{1-u_1}-u_2^{n-1}\, \frac{1- (u_1/u_2)^{n}}{1-u_1/u_2}\right]
\nonumber \\
&=& \langle R_n \rangle +2 \int_q^1 \frac{du_2}{1-u_2} \int_q^{u_2} du_1
\left[\frac{1-u_1^n}{1-u_1}\right]- 2 \int_q^1 \frac{du_2}{1-u_2}\, u_2^{n-1}\,
\int_q^{u_2} du_1\, \left[\frac{1- (u_1/u_2)^{n}}{1-u_1/u_2}\right] \nonumber \\
&=& \langle R_n \rangle +T_2 -T_3
\label{S1.9}
\end{eqnarray}

One can further simplify the term $T_2$ in Eq. (\ref{S1.9}) in the following way
\begin{eqnarray}
T_2 &= & 2 \int_q^1 \frac{du_2}{1-u_2} \int_q^{u_2} du_1
\left[\frac{1-u_1^n}{1-u_1}\right] \nonumber \\
&=& 2 \int_q^1 \frac{du_2}{1-u_2}\left[1-u_2^n+u_2^n\right] \int_q^{u_2} du_1
\left[\frac{1-u_1^n}{1-u_1}\right] \nonumber \\
&=& 2 \int_q^1 \frac{du_2}{1-u_2}\,(1-u_2^n)\, \int_q^{u_2} du_1
\left[\frac{1-u_1^n}{1-u_1}\right] + 2 \int_q^1 \frac{du_2}{1-u_2}\, u_2^n
\int_q^{u_2} du_1
\left[\frac{1-u_1^n}{1-u_1}\right] \nonumber \\
&=& T_{21}+ T_{22}
\label{S1.10}
\end{eqnarray}
The term $T_{21}$ can be exactly integrated by a change of variable:
$z_2=\int_q^{u_2} du_1 (1-u_1^n)/(1-u_1)$, yielding
\begin{eqnarray}
T_{21}= 2 \int_q^1 \frac{du_2}{1-u_2}\,(1-u_2^n)\, \int_q^{u_2} du_1
\left[\frac{1-u_1^n}{1-u_1}\right] 
= 2 \int_0^{\langle R_N\rangle} dz_2\, z_2= \langle R_n\rangle^2\, ,
\label{S1.11}
\end{eqnarray}
where we have used the following fact
\begin{eqnarray}
\int_q^1 du_1\, \frac{1-u_1^n}{1-u_1} \int_q^1 du_1\, \sum_{m=0}^{n-1}
u_1^{m} = \sum_{m=1}^n \frac{1-q^m}{m} = \langle R_n\rangle\, .
\label{S1.12}
\end{eqnarray}
In the final line, we have used the result for the mean number of records in Eq. (7) of 
the main text.
Now, we consider the term $T_3$ in Eq. (\ref{S1.9}). Making the change of variable
$u_1=u_2\, u_1'$, we get
\begin{eqnarray}
T_3 &= & 2 \int_q^1 \frac{du_2}{1-u_2}\, u_2^{n-1}\,
\int_q^{u_2} du_1\, \left[\frac{1- (u_1/u_2)^{n}}{1-u_1/u_2}\right]\nonumber \\
&=& 2 \int_q^1 \frac{du_2}{1-u_2}\, u_2^{n}\, \int_{q/u_2}^1 du_1' \left[\frac{1-
(u_1')^{n}}{1-u_1'}\right]
\label{S1.13}
\end{eqnarray}
Putting all the terms together, we finally get a relatively compact expression for
the variance
\begin{eqnarray}
V_n(q) =  \langle R_n^2\rangle- \langle R_n\rangle^2 &= & \langle R_n\rangle +
2\int_q^1
\frac{du_2}{1-u_2}\, u_2^{n} \left[ \int_q^{u_2} du_1\, \frac{1-u_1^n}{1-u_1}-
\int_{q/u_2}^1 du_1\, \frac{1-u_1^n}{1-u_1}\right]\nonumber \\
&=& \langle R_n \rangle + J_n(q)
\label{S1.14}
\end{eqnarray}
Upon changing $u_2\to u$ and $u_1\to v$, we have
\begin{equation}
J_n(q)= 2\int_q^1
\frac{du}{1-u}\, u^{n} \left[ \int_q^{u} dv\, \frac{1-v^n}{1-v}- 
\int_{q/u}^1 dv\, \frac{1-v^n}{1-v}\right]
\label{S1.Jnq}
\end{equation}
and Eq. (\ref{S1.14}) reduces to Eq. (11) of the main text.
Note that the result in Eq. (\ref{S1.14}) is exact for any $n$ and any 
$0\le q\le 1$. 

\vskip 0.3cm

\noindent {\em Asymptotic behavior of the variance for large $n$ and fixed $q$.} To find 
the asymptotic
large $n$ behavior of $V_n(q)$ in Eq. (\ref{S1.14}) for fixed $q$, we can
use the asymptotic behavior of $\langle R_n\rangle$ given in Eq. (\ref{S1.avg4}).
It remains to estimate the large $n$ behavior of $J_n(q)$ in Eq. (\ref{S1.Jnq}).  
We first show that $J_n(q)\to -\pi^2/6$ as $n\to \infty$, for any fixed $0\le q\le 1$.
To demonstrate this, it is first convenient to make a change of variable
$u'=1-u$ and $v'= 1-v$ in Eq. (\ref{S1.Jnq}), which then reads using $p=1-q$
\begin{equation}
J_n(q)= 2\int_0^p
\frac{du'}{u'}\, (1-u')^{n} \left[ \int_{u'}^{p} \frac{dv'}{v'}\, 
\left(1-(1-v')^n\right)-
\int_{0}^{(p-u')/(1-u')} \frac{dv'}{v'}\, \left(1-(1-v')^n\right)\right]\, .
\label{S1.Jnq1}
\end{equation}
Next, we make a rescaling $u'= u/n$ and $v'=v/n$ to rewrite $J_n(q)$ as
\begin{equation}
J_n(q)= 2\int_0^{pn}
\frac{du}{u}\, \left(1-\frac{u}{n}\right)^{n} \left[ \int_{u}^{pn} \frac{dv}{v}\, 
\left(1-\left(1-\frac{v}{n}\right)^n\right)-
\int_{0}^{(pn-u)/(1-u/n)} \frac{dv}{v}\, 
\left(1-\left(1-\frac{v}{n}\right)^n\right)    \right]\, .
\label{S1.Jnq2}
\end{equation}
It is now convenient to take the $n\to \infty$ limit in Eq. (\ref{S1.Jnq2}) for fixed 
$q=1-p$, which
then reduces to a constant independent of $q$ 
\begin{eqnarray}
J_n(q) &\to & -2\int_0^{\infty} \frac{du}{u} e^{-u}\left[\int_u^{\infty} \frac{dv}{v} 
\left(1-e^{-v}\right)-\int_0^{\infty} \frac{dv}{v} \left(1-e^{-v}\right)\right] \nonumber 
\\
&=& -2 \int_0^{\infty} \frac{du}{u} e^{-u} \int_0^{u} \frac{dv}{v} \left(1-e^{-v}\right) 
\, .
\label{S1.Jasymp1}
\end{eqnarray} 
To evaluate this constant, we use the power series expansion, 
\begin{equation}
\frac{1-e^{-v}}{v}= \sum_{k=1}^\infty (-1)^{k-1} \frac{v^{k-1}}{k!}\, .
\label{S1.int1}
\end{equation}
Hence,
\begin{equation}
\int_0^u \frac{dv}{v} \left(1-e^{-v}\right)= \sum_{k=1}^{\infty} \frac{(-1)^{k-1}}{k!}\, 
\frac{u^k}{k}\, .
\label{S1.int2}
\end{equation}
Substituting (\ref{S1.int2}) in Eq. (\ref{S1.Jasymp1}) and carrying out the integral over 
$u$ gives, using the identity $\int_0^{\infty} du\, e^{-u}\, u^{k-1}= \Gamma(k)=(k-1)!$
\begin{equation}
J_n(q)\to -2 \sum_{k=1}^{\infty} 
\frac{(-1)^{k-1}}{k^2}= -\frac{\pi^2}{6}\, .
\label{S1.Jasymp}
\end{equation} 
Hence, using Eqs. (\ref{S1.14}), (\ref{S1.avg4}) and (\ref{S1.Jasymp}), we 
obtain the two leading terms of the variance $V_n(q)$, for large $n$ and fixed 
$q$
\begin{equation}
V_n(q) = \ln(p\, n) + \gamma_E - \frac{\pi^2}{6} + 
O\left(\frac{1}{n}\right)\, .
\label{S1_varasymp}
\end{equation}

\vskip 0.3cm

\noindent {\em Asymptotic behavior of the variance in the random record limit.}
We now analyse the variance $V_n(q)$ in Eq. (\ref{S1.14}) in the `random record'
model, i.e., in the scaling 
limit,
where
$n\to \infty$, $p=1-q\to 0$, with the product $t= p\, n$ fixed. To derive this scaling
behavior, it is convenient to first make a change of variables $u_1=1-v_1$ and
$u_2=1-v_2$ in the integral $J_n(q)$ in Eq. (\ref{S1.14}). This gives
\begin{eqnarray}
J_n(q)&=&  2 \int_0^p \frac{dv_2}{v_2}\, (1-v_2)^n  \times \nonumber \\
&\times& \left[\int_{v_2}^p
\frac{dv_1}{v_1}\, \left(1-(1-v_1)^n\right) - \int_0^{(p-v_2)/(1-v_2)}
\frac{dv_1}{v_1}\, \left(1-(1-v_1)^n\right)\right] \, .
\label{S1.15}
\end{eqnarray}
Next, we rescale $v_1=p\, y_1$ and $v_2= p\, y_2$ and take the scaling limit
$n\to \infty$, $p\to 0$ with the product $t=p\, n$ fixed. In this limit, Eq.
(\ref{S1.15}) reduces to
\begin{eqnarray}
J_n(q=1-p) &\to & 2 \int_0^1 \frac{dy_2}{y_2}\, e^{-t\, y_2}\left[\int_{y_2}^1
\frac{dy_1}{y_1} \left(1-e^{-t\, y_1}\right)- \int_0^{1-y_2} \frac{dy_1}{y_1}
\left(1-e^{-t\, y_1}\right)\right] \nonumber \\
&=& 2 \int_0^1 \frac{dy_2}{y_2}\, e^{-t\, y_2}\left[ \int_{t\,y_2}^t
\frac{dz_1}{z_1} \left(1-e^{-z_1}\right)- \int_0^{t(1-y_2)}
\frac{dz_1}{z_1} \left(1-e^{-z_1}\right)\right] \nonumber \\
&=& 2 \int_0^t \frac{dz_2}{z_2}\, e^{-z_2} \left[\int_0^t \frac{dz_1}{z_1}
\left(1-e^{-z_1}\right)- \int_0^{z_2} \frac{dz_1}{z_1} \left(1-e^{-z_1}\right)
-\int_0^{t-z_2} \frac{dz_1}{z_1} \left(1-e^{-z_1}\right) \right] \nonumber \\
&=&2 \int_0^t \frac{dz}{z}\, e^{-z}\left[ \mu(t)- \mu(z)- \mu(t-z)\right]
\label{S1.16}
\end{eqnarray}
where in the last line, we used the definition $\mu(t)=\int_0^{t} dz (1-e^{-z})/z$ from
Eq. (12) of the main text.
Thus finally, the variance $V_n(q)$ in Eq. (\ref{S1.14}) can be expressed, in the
scaling limit as
\begin{eqnarray}
V_n(q) &\to & \mu(t) + 2\, \int_0^t \frac{dz}{z}\, e^{-z}\left[ \mu(t)-
\mu(z)- \mu(t-z)\right]\, ,
\label{S1.17}
\end{eqnarray}
where $\mu(t)$ is given in Eq. (12) of the main text. This then provides the detailed derivation
of the result stated in Eq. (13) of the main text.

\section{Derivation of the distribution of the record number}

\begin{figure}
\includegraphics[width=0.8\textwidth]{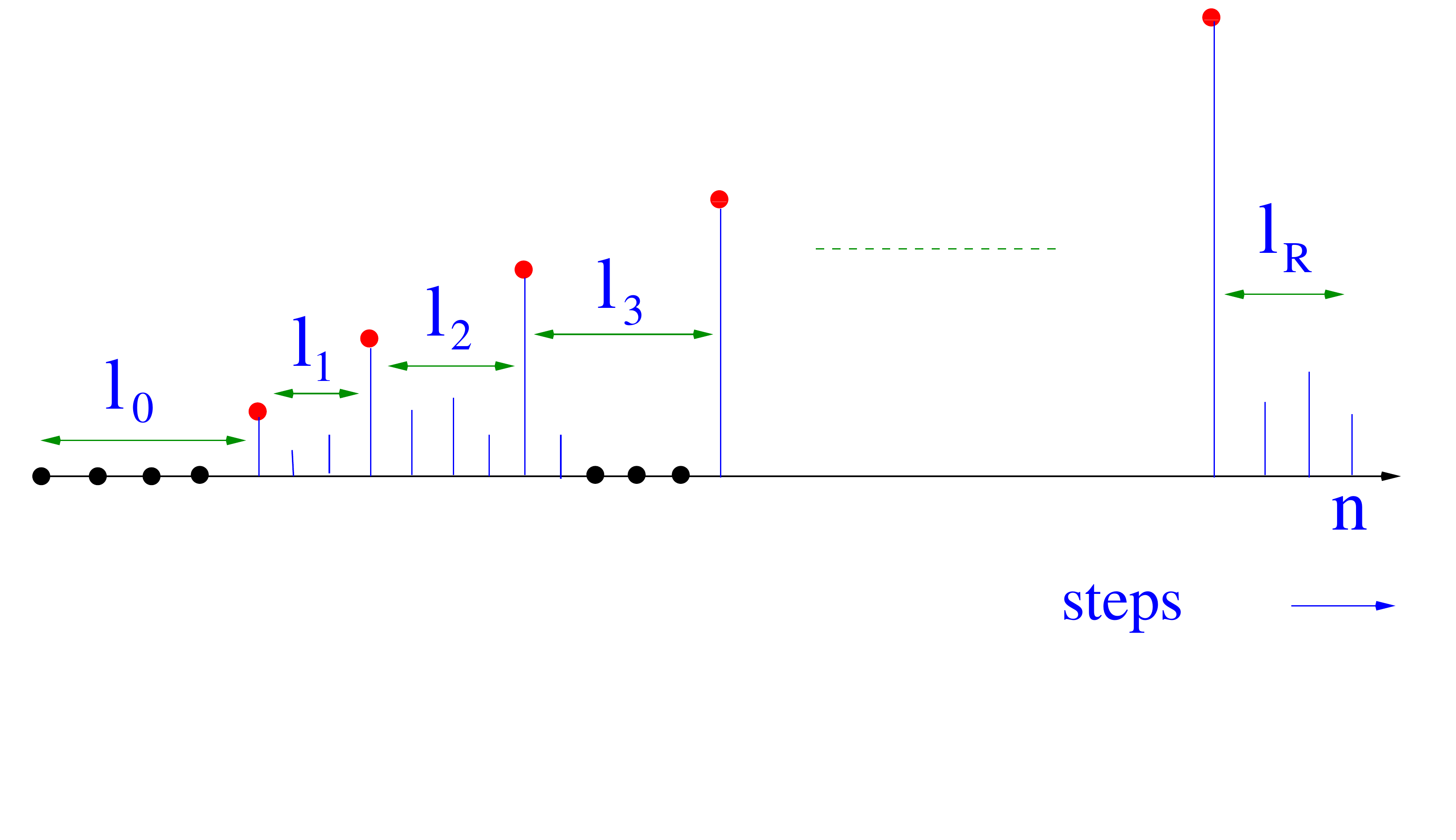}
\caption{A typical configuration of the sequence, with black dots denoting
dry days (no rainfall), blue vertical lines denoting the amount
of rainfall on a `wet' day and the red filled circles (at the top
of a blue vertical line) denoting the record precipitation amounts.
Let $R$ be the number of records in a sequence of $n$ steps.
The sequence $\{l_0,l_1,l_2,\cdots,l_R\}$ denotes the ages of the records.
Note that a record happens necessarily on a wet day.
$l_0$ denotes the number of dry days before the first wet day, hence
the range of $l_0$ is $l_0=0,1,2,\ldots, n$. Similarly, $l_R$ denotes the
age of the last record
before the step $n$ and the range of is $l_R=0,1,2,\ldots, n$.
For all intermediate ages (i.e., excluding $l_0$ and $l_R$), the range is
$l_k=1,2,,\ldots, n$ for
for $k\ne 0$ and $k\ne R$.}
\label{fig.ages}
\end{figure}

In this section we derive the exact double generating function
of the record number distribution quoted in 
Eq. (17) of the main text.
We would like to compute the full distribution of the record number $R_n=\sum_{m=0}^n \sigma_m$ (given in Eq. (3) of the main text), 
i.e., the probability
\begin{equation}
P(R,n)= {\rm Prob.}\left[R_n=R\right].
\label{RN_pdf.1}
\end{equation}
It turns out that, while this representation $R_n= \sum_{m=0}^n \sigma_m$ 
in terms of the binary variables $\sigma_m$'s is useful for
the computation of the mean and variance of $R_n$, it quickly becomes
cumbersome for higher moments. Thus, calculating the full distribution
$P(R,n)$ by this method
seems rather complicated. Hence to compute the full distribution
$P(R,n)$,
we will use a different strategy. It turns out that it is convenient
to consider a more general set of observables, namely
the record number $R$ as well as the set of ages
$\{l_0,l_1,l_2,\cdots, l_R\}$ of the successive records (see Fig.~\ref{fig.ages}). The age $l_k$ of the $k$-th record is the
number of steps between the occurrence of the $k$-th record and
the next $(k+1)$-th record. Note that a record can happen necessarily on a
wet day. We denote by $l_0$ the number of dry days before the first
record, and $l_0$ can take values in the range $l_0=0,1,2,\cdots, n$.
Similarly, $l_R$ denotes the age of the last record till the $n$-th step
and hence $l_R=0,1,2,\cdots, n$. The ages of the intermediate records
(i.e., excluding $l_0$ and $l_R$) can take values, $l_k=1,2,\cdots, n$
for $1\le k\le (R-1)$. Note that the record ages satisfy a sum rule
\begin{equation}
l_0+l_1+l_2+\cdots+ l_R +1 = n\, .
\label{sum_rule.1}
\end{equation}
Our goal is to (i) first write down the
joint distribution of the record number $R$ and the record
ages $\{l_0,l_1,\cdots,l_R\}$ and (ii) then integrate out the record ages
to finally obtain the marginal distribution of the record number $P(R,n)$
only.

To proceed, we define $P(\vec l, R, n)$ as the joint distribution   
of the record ages $\vec l\equiv \{l_0,l_1,l_2\cdots,l_R\}$ and
the record number $R$ in $n$ steps. The marginal distribution
of the record number only, i.e., $P(R,n)$ can then be obtained
from this joint distribution by summing over the record ages
\begin{equation}
P(R,n)= \sum_{\vec l} P(\vec l, R, n)\, .
\label{prn.1}   
\end{equation}
It turns out one can explicitly write down the joint PDF $P(\vec l, R,
n)$ as follows. Let $\{x_1,x_2,\cdots, x_R\}$ denote the precipitation amounts
on the record days, i.e., the record values. Since they are
successive records, we must have $x_1<x_2<x_3<\cdots <x_R$. Also,
these record occurrences must be wet days. Hence we can write 
the joint distribution $P(\vec l, R, n)$ as a nested integral
\begin{eqnarray}
P(\vec l, R,n) &= & \int_{\delta}^{\infty} dx_R\,
\left[(1-q)p_W(x_R)\, \left(\int_0^{x_R}
                    p(x')dx'\right)^{l_R}\right]\, \times
\nonumber \\
&\times &
\int_{\delta}^{x_R} dx_{R-1}\, \left[(1-q)p_W(x_{R-1})\,
\left(\int_0^{x_R}
p(x')dx'\right)^{l_{R-1}-1}\right]\, \cdots \nonumber \\
&\times & \cdots \int_{\delta}^{x_2} dx_1\,
\left[(1-q)p_W(x_1)\,
\left(\int_0^{x_2}
p(x')dx'\right)^{l_{1}-1}\right]\, q^{l_0}\,\, \delta_{l_0+l_1+\ldots+
l_R+1,
n}
\label{age_jpdf.1}
\end{eqnarray}
where $p(x)=q \delta(x)+ (1-q) p_W(x)$ is the effective PDF of 
preciptation given in Eq. (1) of the main text. 
Note again that the lower limit $\delta$ in each
integration refers to the fact that $p_W(x)$ has support only over  
$x\in [\delta,\infty]$. Eventually we take the limit $\delta\to 0$.

The result in Eq. (\ref{age_jpdf.1}) can be understood as follows.
Consider first the value $x_{R}$ of the $R$-th record (the last one)
(see Fig.~\ref{fig.ages}). If $x_R$ is a record,
it has to be a wet day and hence the probability of its occurrence is
$(1-q)\, p_W(x_R)$. Now, given that this is the last record, all $l_R$
days following this must have values less that $x_R$. The probability
of this event is $ \left(\int_0^{x_R} p(x')dx'\right)^{l_R}$, with
$l_R=0,1,2,\cdots\, n$. Hence, the product $\left[(1-q)\, p_W(x_R)
\left(\int_0^{x_R} p(x')dx'\right)^{l_R}\right]$ explains
the first factor in the first line of Eq. (\ref{age_jpdf.1}).
Now consider the last but one record, i.e., $x_{R-1}$.
The probability of its occurrence is again $(1-q)\, p_W(x_{R-1})$
and all the days between the $(R-1)$-th record and the $R$-th record
(and there are $(l_{R-1}-1)$ such days) must have values less than   
$x_{R-1}$ if $x_R$ is a record. Hence, the product $\left[(1-q)\,
p_W(x_{R-1})
\left(\int_0^{x_{R-1}} p(x')dx'\right)^{l_{R-1}-1}\right]$ explains
the second factor in the first line of Eq. (\ref{age_jpdf.1}).
Similarly one can proceed in a nested way. Finally, one needs to
integrate over the record values $\{x_1,x_2,\cdots, x_R\}$, but
respecting the constraint $x_1<x_2<x_3<\cdots<x_R$. This explains the
limits of the integrations.
The last factor $q^{l_0}$ denotes the
probability that there are exactly $l_0$ dry days (each occurs with
probability $q$ independently) before the first wet day occurs (the first
wet day is
necessarily a record day with value $x_1$). Finally, the record ages
must satisfy the sum rule in Eq. (\ref{sum_rule.1}), explaining 
the Kronecker delta function in Eq. (\ref{age_jpdf.1}).

To proceed, we first make the customary change of variables as in Eq. 
(\ref{S1.4}), namely
\begin{equation}
u= \int_0^x p(x') dx' = q\, \theta(x) + (1-q) \int_\delta^x p_W(x')\,
dx'\, .
\label{cov.1}
\end{equation}
With this change of variable and taking $\delta\to 0^+$ limit, the 
explicit dependence on $p_W(x)$
disappears and Eq. (\ref{age_jpdf.1}) transforms into
\begin{equation}
P(\vec l, R,n) =  q^{l_0}\, \int_q^{1} du_k\, u_R^{l_R}
\int_q^{u_R} du_{R-1}\, u_{R-1}^{l_{R-1}-1}\cdots \int_q^{u_2} du_1\,
u_1^{l_1-1} \,\, \delta_{l_0+l_1+\ldots+
l_R+1,n}
\label{age_jpdf.2}
\end{equation}
Now, to get rid of the delta function constraint, we consider the
generating function, i.e., we multiply both sides of Eq.
(\ref{age_jpdf.2}) by $z^n$ and sum over $n$, as well as over $\vec l$.
When we sum over $\vec l$, we recall that while $l_0=0,1,2\cdots,$ and
$l_R=0,1,2\cdots$, all other $l_k=1,2,3,\cdots$ (for $k\ne 0$ and $k\ne
R$). This gives
\begin{equation}
\sum_{\vec l} \sum_{n=1}^{\infty} P(\vec l, R, n)\, z^n= 
\frac{1}{1-q\,z}\, \int_q^{1}
\frac{z\, du_R}{1-u_R\, z}\,\int_q^{u_R} \frac{z\, du_{R-1}}{1-
u_{R-1}\,z}\cdots \int_q^{u_2} \frac{z\, du_1}{1-u_1\, z}\, .
\label{age_jpdf.3}
\end{equation}
This can be further simplified by making the change of variables, $u_k
z=v_k$, to give
\begin{equation}
\sum_{\vec l} \sum_{n=1}^{\infty} P(\vec l, R, n)\, z^n= 
\frac{1}{1-q\,z}\,\int_{qz}^z
\frac{dv_R}{1-v_R}\, \int_{qz}^{v_R} \frac{dv_{R-1}}{1- v_{R-1}}\cdots
\int_{qz}^{v_2} \frac{dv_1}{1-v_1}\, .
\label{age_jpdf.4}
\end{equation}
This last nested integral can be computed explicitly as follows. Let us
first rewrite Eq. (\ref{age_jpdf.4}) as
\begin{equation}
\sum_{\vec l} \sum_n P(\vec l, R, n)\, z^n= \frac{1}{1-q\,z}\, W_R(z,z)
\label{age_jpdf.5}
\end{equation}
where we define the following nested integral
\begin{equation}
W_R(x,z)= \int_{qz}^{x} \frac{dv_R}{1-v_R}\, \int_{qz}^{v_R}
\frac{dv_{R-1}}{1- v_{R-1}}\cdots
\int_{qz}^{v_2} \frac{dv_1}{1-v_1}\, .
\label{nested.1}
\end{equation} 

To evalute $W_R(x,z)$,
we take the derivative of Eq. (\ref{nested.1}) with respect to $x$ for fixed 
$z$. For simplicity of notation, we denote this derivative by an ordinary 
derivative and not a partial derivative ($z$ can be thought of just a
parameter in $W_R(x,z)$). We find 
that $W_R(x,z)$ satisfies the recursion relation
\begin{equation}
\frac{dW_R(x,z)}{dx}= \frac{1}{1-x}\, W_{R-1}(x,z)\, ; \quad\quad {\rm
for}\,\,\,\, R\ge 2
\label{recur.1} 
\end{equation} 
starting from
\begin{equation}
W_1(x,z)= \int_{qz}^{x} \frac{dv}{1-v}= 
-\ln\left(\frac{1-x}{1-qz}\right)\, .
\label{recur.2}
\end{equation}
We can now check easily that the solution of the recursion relation
(\ref{recur.1}), satisfying the initial condition in (\ref{recur.2}) is
given by
\begin{equation}
W_R(x,z)= \frac{1}{R!}\, 
\left[-\ln\left(\frac{1-x}{1-qz}\right)\right]^R\, .
\label{recur_sol.1}
\end{equation}
Substituting this result (\ref{recur_sol.1}) for $W_R(x=z,z)$ in Eq.
(\ref{age_jpdf.5}), we obtain our final result
\begin{equation}
\sum_{n=1}^{\infty} P(R,n)\, z^n= \sum_{\vec l} \sum_{n=1}^{\infty} P(\vec 
l, R, n)\, z^n=
\frac{W_R(z,z)}{1-q\,z}= \frac{1}{1-q\,z}\, \frac{1}{R!}\, \left[
-\ln\left(\frac{1-z}{1-qz}\right)\right]^R\, ; \quad\quad R\ge 1 \, .
\label{dist_sol.1}
\end{equation}
For $R=0$, we have $P(0,n)=q^n$ since the probability of having no records is
the same as the probability that all $n$ days are dry. Hence,
\begin{equation}
\sum_{n=0}^{\infty} P(0,n)\, z^n= \frac{1}{1-qz}\, ; \quad\quad R=0 \, .
\label{dist_solR0}
\end{equation}
As a nontrivial check one can verify that $P(R,n)$ is normalized to unity. 
Summing Eq. (\ref{dist_sol.1})
over all $R=1,2\ldots$ and Eq. (\ref{dist_solR0}) for $R=0$, one
obtains (using $P(R,0)=\delta_{R,0}$ and a few minor steps of algebra)
\begin{equation}
\sum_{n=0}^{\infty} \sum_{R=0}^\infty  P(R, n)\, z^n= \frac{1}{1-z}\,
\label{norm.1}
\end{equation}
indicating that $\sum_{R=0}^{\infty} P(R,n)=1$.
Furthermore, by taking the
derivative of Eq. (\ref{dist_sol.1}) with respect to $z$ and setting
$z=1$, one can show that one recovers the result for the mean given in Eq.
(7) of the main text. Similarly, taking derivatives twice with respect to 
$z$ and setting $z=1$, one recovers, after straightforward algebra,
the result for the second moment in Eq. (11) of the main text, obtained
by a different method (using correlations of the $\sigma_m$'s).

Furthermore, multiplying Eq. (\ref{dist_sol.1}) by $\lambda^R$ and summing over $R=1,2,3, 
\ldots$, one gets
\begin{eqnarray}
\sum_{n=1}^{\infty} \sum_{R=1}^{\infty} P(R,n) \lambda^R\, z^n &= &
\frac{1}{1-qz}\, \sum_{R=1}^{\infty} \frac{{\lambda}^R}{R!}\, \left[
-\ln\left(\frac{1-z}{1-qz}\right)\right]^R \nonumber \\ 
&=& \frac{(1-qz)^{\lambda-1}}{(1-z)^{\lambda}}- \frac{1}{1-qz}\, .
\label{dgf_iid.0}
\end{eqnarray}
Including the terms corresponding to $n=0$ and $R=0$ (using $P(0,n)=q^n$ and
$P(R,0)=\delta_{R,0}$) on the left hand side of Eq. (\ref{dgf_iid.0}), we
can finally write a compact expression for the double generating function
\begin{equation}
\sum_{n=0}^{\infty}\sum_{R=0}^{\infty} P(R,n) \lambda^R\, z^n = 
\frac{(1-qz)^{\lambda-1}}{(1-z)^{\lambda}}\, .
\label{dgf_iid.1}
\end{equation}
This completes the derivation of Eq. (17) in the main text.

Note that for $q=0$, Eq. (\ref{dgf_iid.1}) reduces to
\begin{equation}
\sum_{n=0}^{\infty} \sum_{R=0}^{\infty} P_{q=0}(R,n)\,\lambda^R\,  z^n= 
(1-z)^{-\lambda}\, . 
\label{dist_iid.1}
\end{equation}
This result for $q=0$ was already known in the literature in a slightly 
different 
disguise. In fact, it is well known that the number of records $R$ of $n$ 
independent and identically (and continuously) distributed (i.i.c.d.)  
variables has the same statistical law as the number of cycles $R$ in 
a random permutation of $n$ 
elements~\cite{Renyi,Goldie,Flajolet-Sedgewick}. This connection has also
appeared in various statistical physics problems, such as in 
growth processes on networks~\cite{GL2008} and in a class of one
dimensional ballistic aggregation models~\cite{MMS2009}. The double generating 
function for the distribution of the number 
of cycles in random permutation of $n$ elements was known to
have the form in Eq. (\ref{dist_iid.1}) with $R$ denoting the
number of cycles. Thus our result for arbitrary $0\le q\le 1$ in Eq. 
(\ref{dgf_iid.1}) provides a 
generalization of the $q=0$ result in Eq. (\ref{dist_iid.1}). 
There is a precise combinatorial interpretation of our formula
for general $0\le q\le 1$ in terms of the number of cycles in a random permutation
of $n$ elements. Indeed, consider a dilute version of the permutation problem, where each 
of the $n$ elements is either present with probability $p=1-q$, or absent with
probability $q$. Then, the number of `present' elements becomes a random variable 
with binomial distribution, and consequently the number of cycles $R$ of the
random permutation of the `present' elements is precisely our $P(R,n)$ with
a general binomial parameter $p=1-q$ (see also Sect.~\ref{Sec:Alternative}).

\begin{figure}
\includegraphics[width=0.8\textwidth]{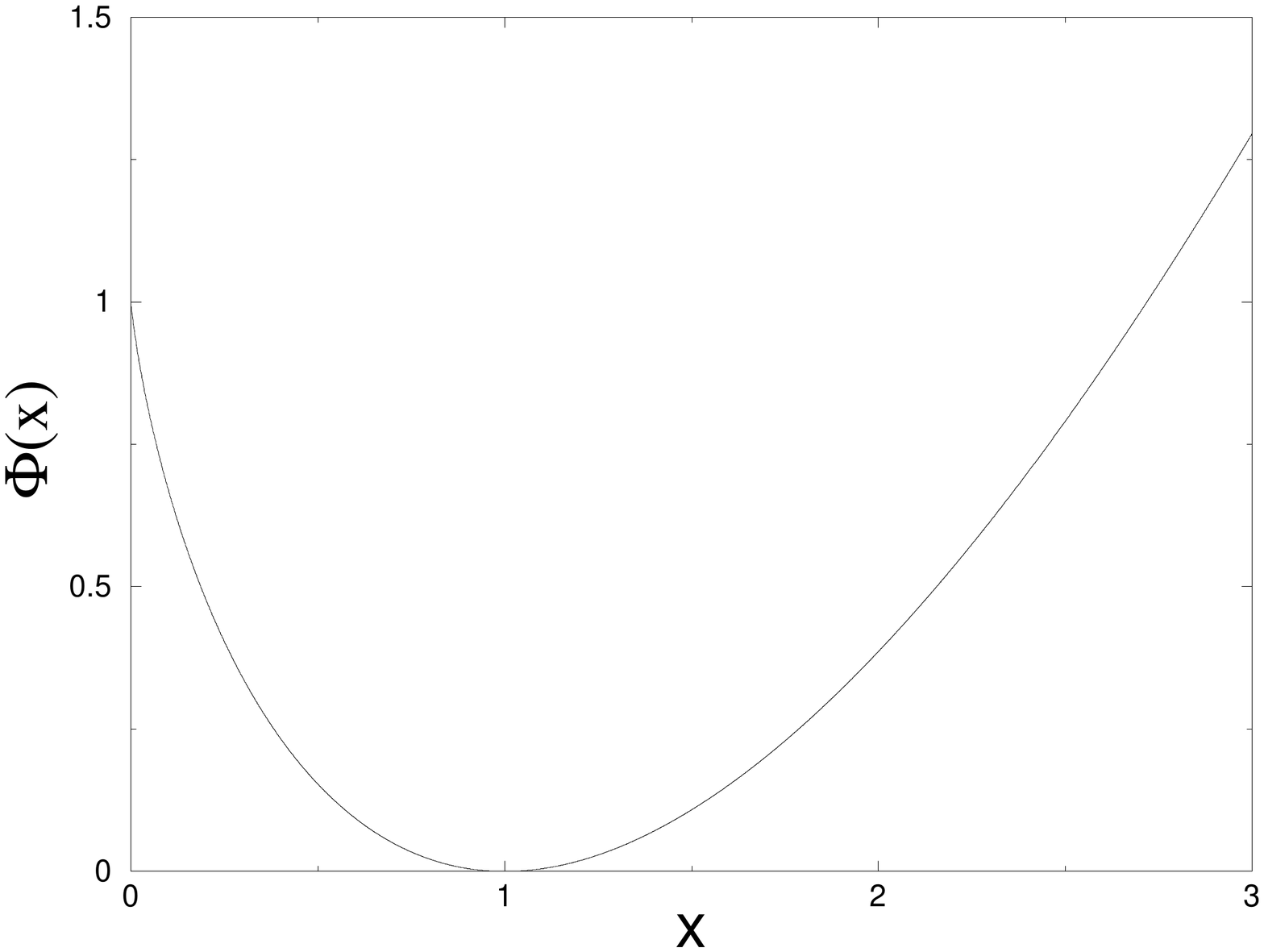}
\caption{The large deviation rate function $\Phi(x)=1-x-x\, \ln x$ plotted
as a function of $x$. The rate function has a unique minimum at $x=x^*=1$ around 
which it has a quadratic behavior, $\Phi(x)\approx (x-1)^2/2$.}
\label{fig.ldf1}
\end{figure}

\section{Asymptotic behavior of $P(R,n)$ for large $n$}

In this section, we perform an asymptotic analysis of the double 
generating function in Eq. (\ref{dgf_iid.1}) to derive 
the large $n$ behavior of $P(R,n)$, for general $q$. 
To proceed, it is first convenient to set
$z=e^{-\mu}$ in Eq. (\ref{dgf_iid.1}). Now, for large $N$, the most important 
contribution comes from the vicinity of $\mu=0$ (or $z=1$). Expanding the
r.h.s. of Eq. (\ref{dgf_iid.1}) for small $\mu$, one gets, to leading 
order in 
$\mu$,
\begin{equation}
\sum_{n=1}^{\infty}\sum_{R=1}^{\infty} P(R,n) \lambda^R\, e^{-\mu\, n}\approx 
\frac{p^{\lambda-1}}{\mu^{\lambda}}\, ,
\label{S3.1}
\end{equation}
where we used $q=1-p$. Note that since, to this leading order, the contribution
from the pole at $z=1/q$ in Eq. (\ref{dgf_iid.1}) is neglected, we do not
include the $R=0$ term in the sum on the left hand side of Eq. (\ref{dgf_iid.1}). 
Using 
the identity, $\int_0^{\infty} n^{\lambda-1} e^{-\mu n} dn=  
\Gamma(\lambda)\, \mu^{-\lambda}$, 
one can then invert the Laplace
transform with respect to $n$ in Eq. (\ref{S3.1}). This gives, for large 
$n$,
\begin{eqnarray}
\sum_{R=1}^{\infty} P(R,n) \lambda^R &\approx & \frac{(p\, 
n)^{\lambda-1}}{\Gamma(\lambda)} \nonumber \\
& = & \frac{1}{pn}\, \frac{\lambda\, (pn)^{\lambda}}{\Gamma(1+\lambda)}\, ,
\label{S3.2}
\end{eqnarray}
where we used the identity $\Gamma(\lambda)= \Gamma(1+\lambda)/\lambda$.
The next step is to expand the right hand side (rhs) of Eq. (\ref{S3.2}) in a power series in
$\lambda$ and identify the coefficient of $\lambda^R$. 
For this, we use 
\begin{equation}
(pn)^{\lambda}= \sum_{k=0}^{\infty} \frac{\left(\ln (pn)\right)^{k}}{k!}\, \lambda^k
\label{S3.p1}
\end{equation}
and the power series expansion 
\begin{equation}
\frac{1}{\Gamma(1+\lambda)}= \sum_{m=0}^{\infty} d_m\, \lambda^m
\label{S3.p2}
\end{equation}
where $d_0=1$. Expanding the rhs of Eq. (\ref{S3.2}) using (\ref{S3.p1}) and 
(\ref{S3.p2}) and identifying the power of $\lambda^R$ gives, for fixed $R\ge 1$
\begin{equation}
P(R,n)\approx \frac{1}{pn} \sum_{m=0}^{R-1} \frac{\left(\ln 
(pn)\right)^{R-1-m}}{(R-1-m)!}\, d_m \,.
\label{S3.p3}
\end{equation}
Finally, noticing that for large $n$, the dominant contribution comes from the $m=0$
term in the rhs of Eq. (\ref{S3.p3}), we get for large $n$ and fixed $R\ge 1$
\begin{equation}
P(R,n) \approx \frac{1}{pn}\, \frac{ \left(\ln (pn)\right)^{R-1}}{(R-1)!} 
\label{S3.poisson}
\end{equation}
which is just a Poisson distribution with parameter $\ln (pn)$.
This provides the derivation of Eq. (18) of the main text. Finally, in the
limit when both $R\to \infty$ and $\ln(pn)\to \infty$, but with the ratio
$x= R/\ln(pn)$ fixed, we can use Stirling formula to express the rhs
of Eq. (\ref{S3.poisson}) in a large deviation form
\begin{equation}
P(R,n) \sim e^{-\ln (p\, n)\, \Phi(x)}= e^{-\ln (p\, n)\, 
\Phi\left(\frac{R}{\ln (p\, n)}\right)}\, 
\label{S3.7}
\end{equation}
where the rate function $\Phi(x)$ is given by
\begin{equation}
\Phi(x)= 1-x+ x\, \ln x 
\, ,
\label{S3.8}
\end{equation}
as reported in Eqs.~(19) and (20) of the main text.
Interestingly, the rate function $\Phi(x)$ is independent of $q$. The $q$
dependence appears only in renormalizing $n$ to $p\,n = (1-q)\, n$.
Indeed, even for the case $q=0$ (i.i.c.d.), we are not aware of any result
in the literature pointing out this explicit large deviation form.
The rate function, plotted in Fig.~\ref{fig.ldf1}, has a unique minimum
at $x^*=1$, where it has a quadratic behavior, $\Phi(x)\approx (x-1)^2/2$.
This means, from Eq. (\ref{S3.7}), that $P(R,n)$ is maximal near $x=1$, i.e.,
at $R=\ln (p\, n)$. Indeed, using the quadratic behavior near $x=1$, we see that
the typical fluctuations of $R$ are described by a Gaussian form
\begin{equation}
P(R,n) \approx \frac{1}{\sqrt{2\pi \ln(p\,n)}}\,
e^{- \frac{(R-\ln(p\, n))^2}{2\ln(p\, n)}}
\label{S3.10}
\end{equation}
with mean $\ln (p\, n)$ and variance $\ln (p\,n)$. 
% This gives the derivation of the result in Eq. (20) in the main text.

\section{An alternative derivation of Eq. (\ref{dgf_iid.1})}
\label{Sec:Alternative}

There is an alternative way to compute the distribution of the record 
number $P(R,n)$ in the Bernoulli model for arbitrary $0\le q\le 1$ ($q$ 
being the probability
that a dry day occurs), knowing already the result for the $q=0$ case.
Consider a sequence of $n$ days, and let $n_W$ denote the number of wet 
days, while $n-n_W$ denotes the number of dry days. Given that a wet day 
occurs with probability $p=1-q$, it follows that the number of wet days
$n_W$ has a binomial distribution
\begin{equation}
Q(n_W,n)= \binom{n}{n_W}\, p^{n_W}\, q^{n-n_W}\, ;\quad\quad {\rm 
where}\quad n_W=0,1,2,\ldots, n.
\label{S4.1}
\end{equation}
Now, a record can happen only on a wet day. Let ${\rm Prob}(R, n_W)$
denote the probability of having $R$ records among $n_W$ wet days.
Thus, ${\rm Prob}(R, n_W)$ is just the record number distribution of the 
pure i.i.c.d. case (i.e., $q=0$). Hence we have 
\begin{equation}
{\rm Prob}(R, n_W)= P_{q=0}(R, n_W)\, 
\label{S4.2}
\end{equation}
where the double generating function of $P_{q=0}(R, n_W)$ satisfies
Eq. (\ref{dist_iid.1}), i.e.,
\begin{equation}
\sum_{n_W=0}^{\infty} \sum_{R=0}^{\infty} P_{q=0}(R,n_W)\, 
\lambda^R\, u^{n_W}= 
(1-u)^{-\lambda}\, .
\label{dist_iid.repeat}
\end{equation}
Now, knowing $P_{q=0}(R, n_W)$, it is clear that $P(R,n)$ for 
fixed $n$ and arbitrary 
$q$ can be written simply as
\begin{eqnarray}
P(R, n)&=& \sum_{n_W=0}^n {\rm Prob}(R, n_W)\, Q(n_W,n)\nonumber \\
&=& \sum_{n_W=0}^n P_{q=0}(R, n_W)\, \binom{n}{n_W}\, p^{n_W}\, 
q^{n-n_W}\, .
\label{S4.3}
\end{eqnarray}
Thus, basically it amounts to studying the record number distribution 
of just the i.i.c.d. case,  albeit
with a random number of $n_W$ entries and one needs to average over $n_W$.

To compute the double generating function of $P(R,n)$ using the
exact formula in Eq (\ref{S4.3}), it is useful to first formally invert
Eq. (\ref{dist_iid.repeat}) with respect to $u$ using Cauchy's theorem.
This gives
\begin{equation}
\sum_{R=0}^{\infty} P_{q=0}(R,n_W)\,
\lambda^R = \int_{C_0} \frac{du}{2\pi\, i}\, \frac{1}{u^{n_W+1}}\, 
(1-u)^{-\lambda}\,
\label{S4.4}
\end{equation}
where $C_0$ is any contour encircling the origin in the complex $u$ plane.
Now, multiplying Eq. (\ref{S4.4}) by the binomial distribution $Q(n_W,n)$
in Eq. (\ref{S4.1}) and summing over $n_W$, we get
\begin{eqnarray}
\sum_{n_W=0}^n Q(n_W,n) \sum_{R=0}^{\infty} P_{q=0}(R,n_W)\,
\lambda^R &= & \int_{C_0} \frac{du}{2\pi\, i}\, \frac{1}{u}\, 
(1-u)^{-\lambda}\, 
\sum_{n_W=0}^n
\binom{n}{n_W}\, \left(\frac{p}{u}\right)^{n_W}\, q^{n-n_W} 
\nonumber \\
& =& \int_{C_0} \frac{du}{2\pi\, i}\, \frac{1}{u}\, (1-u)^{-\lambda}\, 
\left(\frac{p}{u}+q\right)^n\, .
\label{S4.5}
\end{eqnarray}
We next multiply Eq. (\ref{S4.5}) by $z^n$ and sum over $n$.
To ensure the convergence of the geometric series, we need to
assume $u >pz/(1-qz)$ for a given $z$. Indeed, we can do this 
by deforming the original contour $C_0$, such that it includes
$u=pz/(1-qz)$ inside it. Once ensured of the convergence,
summing over $n$ we 
get, upon using  
Eq. (\ref{S4.3}), the following identity
\begin{eqnarray}
\sum_{n=0}^{\infty} \sum_{R=0}^{\infty}
P(R,n) \lambda^R\,z^n
&=& \int_{C_0} \frac{du}{2\pi\, 
i}\, \frac{1}{u}\, 
(1-u)^{-\lambda}\, \frac{1}{1- 
z\left(\frac{p}{u}+q\right)} \nonumber \\
&=& \frac{1}{1-qz}\, \int_{C_0} \frac{du}{2\pi\,
i}\,
(1-u)^{-\lambda}\, \frac{1}{u- \frac{pz}{1-qz}}\, .
\label{S4.6}
\end{eqnarray}
Finally, noting that there is a simple pole at $u= pz/(1-qz)$,
and since our deformed contour $C_0$ contains this pole inside it, the
integral is just given by the residue at the pole $u=pz/(1-qz)$. This 
gives, using $p+q=1$, the desired result
\begin{equation}
\sum_{n=0}^{\infty}\sum_{R=0}^{\infty} P(R,n) \lambda^R\, 
z^n=\frac{(1-qz)^{\lambda-1}}{(1-z)^{\lambda}}\, 
\label{S4.7}
\end{equation}
which was derived before in Eq. (\ref{dgf_iid.1}) using a completely 
different method exploiting the renewal structure of the underlying record 
process.

\end{document}